\documentclass{article} 
\usepackage{iclr2026_conference,times}

\usepackage{microtype}
\usepackage{graphicx}
\usepackage{subcaption}
\usepackage{booktabs}
\usepackage{multirow}
\usepackage{amsmath,amssymb,amsthm}
\usepackage{algorithm}
\usepackage{algorithmic}
\usepackage{hyperref}
\usepackage{url}
\usepackage{natbib}
\usepackage{xcolor}
\usepackage{tabularx}

\newcommand{\CASCADE}{\textsc{CASCADE }}

\newcommand{\placeholderfig}[4][]{%
  \begin{center}
  \IfFileExists{#2}{\includegraphics[#1]{#2}}{%
    \fbox{\parbox[c][0.22\textheight][c]{0.92\linewidth}{\centering \texttt{\detokenize{#2}}\\ \vspace{2mm} #3}}%
  }
  \end{center}
  \vspace{-2mm}
  {\footnotesize #4}
}

\title{\CASCADE: Cascaded Scoped Communication for Multi-Agent Re-planning in Disrupted Industrial Environments}

\author{Mingjie Bi \\
State Key Laboratory of General Artificial Intelligence, \\ Beijing Institute for General Artificial Intelligence (BIGAI)\\
Beijing, China \\
\texttt{bimingjie@bigai.ai} \\
}

\newcommand{\AIMSfig}[2]{%
  \IfFileExists{#1}{\includegraphics[width=\linewidth]{#1}}{%
    \fbox{\parbox[c][0.22\textheight][c]{0.98\linewidth}{\centering
      \small\textbf{Figure placeholder}\\[0.25em]
      \texttt{#1}\\[0.5em]
      \footnotesize #2
    }}%
  }%
}
\iclrfinalcopy 
\begin{document}

\maketitle

\begin{abstract}
Industrial disruption replanning demands multi-agent coordination under strict latency and communication budgets, where disruptions propagate through tightly coupled physical dependencies and rapidly invalidate baseline schedules and commitments. Existing coordination schemes often treat communication as either effectively free (broadcast-style escalation) or fixed in advance (hand-tuned neighborhoods), both of which are brittle once the disruption footprint extends beyond a local region. We present \CASCADE, a budgeted replanning mechanism that makes communication scope explicit and auditable rather than fixed or implicit. Each agent maintains an explicit knowledge base, solves role-conditioned local decision problems to revise commitments, and coordinates through lightweight contract primitives whose footprint expands only when local validation indicates that the current scope is insufficient. This design separates a unified agent substrate (Knowledge Base / Decision Manager / Communication Manager) from a scoped interaction layer that controls who is contacted, how far coordination propagates, and when escalation is triggered under explicit budgets. We evaluate \CASCADE on disrupted manufacturing and supply-chain settings using unified diagnostics intended to test a mechanism-design claim---whether explicit scope control yields useful quality--latency--communication trade-offs and improved robustness under uncertainty---rather than to provide a complete algorithmic ranking.
\end{abstract}

\section{Introduction}
Disruptions are routine in industrial systems~\citep{fu2025digital}.
Machine breakdowns, capacity losses, lead-time inflation, and demand shocks can invalidate pre-planned schedules, and material flows within minutes~\citep{tang2006scrm}.
Because these effects propagate through tightly coupled temporal, capacity, and flow dependencies, effective response requires \emph{distributed} replanning across heterogeneous entities---resources, process stages, suppliers, carriers, and distributors.
Even localized shocks can induce system-wide losses through networked dependencies, motivating resilience-oriented decision making under time pressure~\citep{snyder2016orms,ivanov2019disruption}.
A central challenge is coordination under limited communication.
Broadcast-style negotiation may recover feasibility, but it floods the network and consumes scarce response time.
Strictly local coordination is cheap, but it can fail silently when disruption effects cross neighborhood boundaries.
What is missing is a replanning mechanism that explicitly controls the communication footprint: \emph{who} is involved, \emph{how far} coordination propagates, and \emph{when} escalation is warranted.

\textbf{Industrial control context and positioning.}
Distributed scheduling and manufacturing control must balance solution quality against negotiation and information-exchange overhead~\citep{fu2021dist_sched}.
More broadly, recent work in cyber-physical manufacturing and industrial multi-agent control has moved toward modular learning-and-control architectures, dynamic resource reallocation, and resilient distributed decision-making under disruptions~\citep{kovalenko2022toward,bi2021dynamic}.
Related supply-chain studies have likewise developed model-based and network-aware multi-agent approaches for agile disruption response~\citep{bi2022model}.
These works establish the need for decentralized commitment revision in industrial environments, but they still leave the communication scope largely as an implementation choice rather than an explicit mechanism variable.

\textbf{Communication-constrained coordination.}
Related communities increasingly treat communication as a limited decision resource rather than a free side channel.
In communication-constrained MARL, recent methods learn when and how to exchange compressed or context-dependent messages under bandwidth limits, as illustrated by context-aware protocols such as CACOM~\citep{li2024_cacom}.
In parallel, task-oriented communication studies optimize messages for downstream control utility under explicit bit budgets, rather than for faithful reconstruction, including task-oriented data compression for multi-agent coordination~\citep{mostaani2024_tocd_scale}.
These lines strongly motivate explicit, utility-aware communication, but they are rarely studied in disruption-centered industrial replanning, where commitments, feasibility, and escalation budgets are tightly coupled.

Taken together, the literature points to a missing mechanism layer for industrial disruption response:
communication should be \emph{explicitly budgeted and interpretable}, its \emph{scope should adapt} to the evolving disruption footprint, and evaluation should expose systematic \emph{quality--latency--communication} trade-offs under tight operational budgets.
However, industrial disruption replanning differs from standard scheduling or generic distributed optimization in two under-addressed ways:
(i) feasibility depends on role-dependent commitments under coupled temporal, capacity, and flow constraints, often under uncertainty; and
(ii) communication is itself operationally budgeted, so naive escalation can dominate response time.
This leaves a core question:
\emph{how should disruption replanning regulate communication scope as an explicit control variable while still making progress when local repair fails?}

\textbf{Contributions.}
\CASCADE answers this question by treating communication scope as a decision variable.
Agents attempt local resolution first and expand coordination only when validation gates indicate that the current scope is insufficient.
The contributions are:
(i) a disruption-replanning formulation over coupled physical and communication graphs under explicit communication and time budgets;
(ii) \CASCADE{} itself---a unified agent substrate (Knowledge Base / Decision Manager / Communication Manager) together with contract-style negotiation and gate-triggered scoped propagation; and
(iii) a unified diagnostic view that exposes quality--latency--communication trade-offs, propagation footprint, and robustness across disrupted manufacturing and supply-chain environments.

\begin{figure}[t]
\centering
\includegraphics[width=0.85\linewidth]{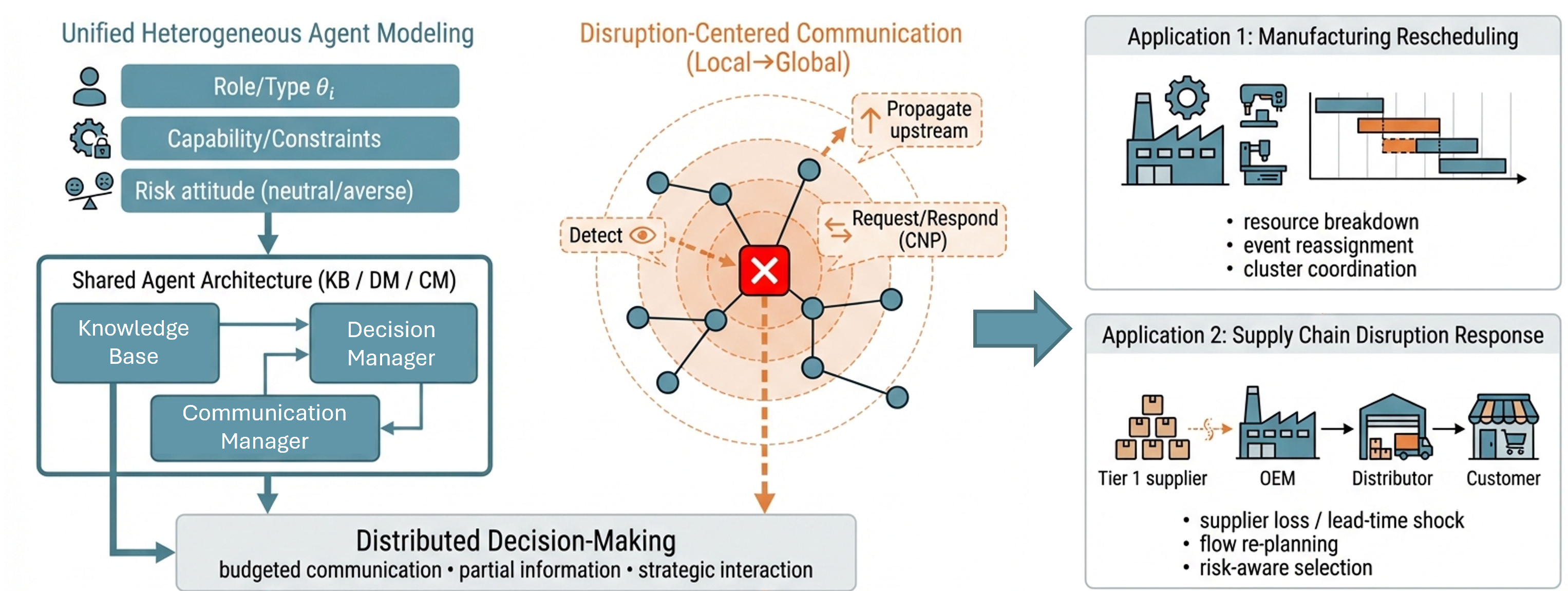}
\caption{A unified agent substrate (KB/DM/CM) supports heterogeneous local modeling, while gate-triggered scoped communication escalates from local to broader coordination only when needed.}
\label{fig:framework}
\end{figure}

\section{Problem Setting}
A disrupted industrial environment is represented by two coupled graphs.
The \emph{physical dependency graph} $G_p=(V,E_p)$ captures material, operation, and flow dependencies.
Each node $i\in V$ is an entity such as a resource, process stage, supplier, transporter, or distributor, and each typed edge $(i\!\rightarrow\! j,k)\in E_p$ denotes a commitment to provide item/type $k$ with quantity and timing constraints.
In parallel, agents coordinate over an \emph{agent communication graph} $G_a=(A,L)$, where $A$ indexes decision-making entities and $(a\!\leftrightarrow\! b)\in L$ indicates that $a$ and $b$ can exchange messages directly.

A baseline plan $\pi_0$ encodes local decisions and cross-agent commitments.
In unified form, a plan $\pi$ can be written as
\begin{equation}
    \pi=\{(c_{ij}^k,\tau_{ij}^k)\}_{(i\!\rightarrow\! j,k)\in E_p}\cup \{u_a\}_{a\in A},
\end{equation}

where $(c_{ij}^k,\tau_{ij}^k)$ are committed transfers and $u_a$ denotes agent-local allocations.
A disruption $\delta$ modifies local feasibility by changing capacities, lead times, costs, or demands, i.e., $\mathcal{C}_a \leftarrow \mathcal{C}_a(\delta)$ for affected agents, and may induce unmet needs that propagate through $G_p$.

The goal is to compute a revised plan $\pi$ that restores feasibility while balancing service loss, operational cost, plan deviation, and communication:
\begin{equation}
    J(\pi)=J_{\mathrm{service}}(\pi)+\lambda_c J_{\mathrm{cost}}(\pi)+\lambda_\Delta d(\pi,\pi_0)+\lambda_{\mathrm{msg}} M(\pi),
\end{equation}

where $d(\pi,\pi_0)$ measures deviation from the baseline plan and $M(\pi)$ measures communication on $G_a$.
Replanning is subject to hard budgets: message $B_{\mathrm{msg}}$ or round limits $B_{\mathrm{round}}$ and a wall-clock deadline $B_{\mathrm{time}}$.
The resulting mechanism problem is therefore not only to restore feasibility, but to do so while explicitly regulating communication footprint under operational constraints.

\section{\CASCADE: Architecture and Mechanism}
\begin{figure}[t]
\centering
\includegraphics[width=\linewidth]{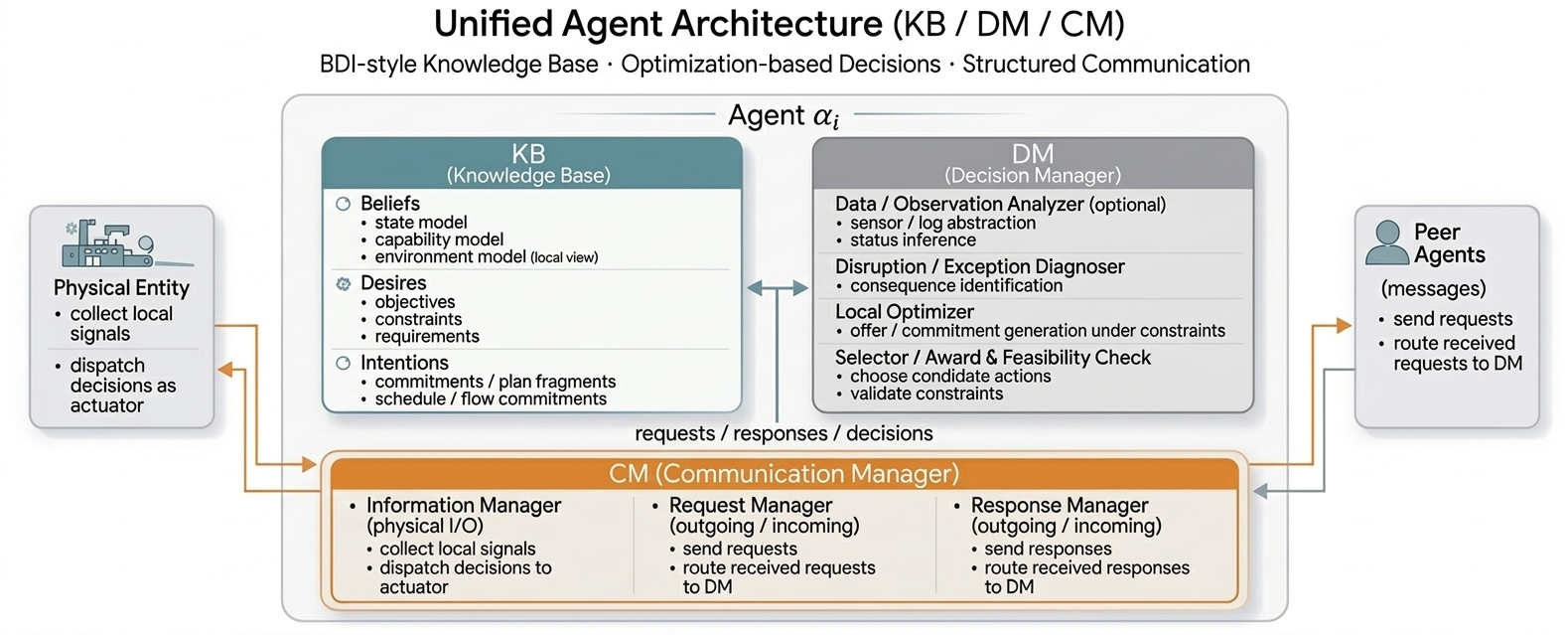}
\caption{\textbf{Unified agent architecture (KB/DM/CM).}
KB maintains local beliefs, priorities, intentions, and uncertainty; DM diagnoses disruptions and computes local commitments; CM routes physical I/O and peer messages.}
\label{fig:agent}
\end{figure}

\begin{figure}[t]
\centering
\includegraphics[width=0.9\linewidth]{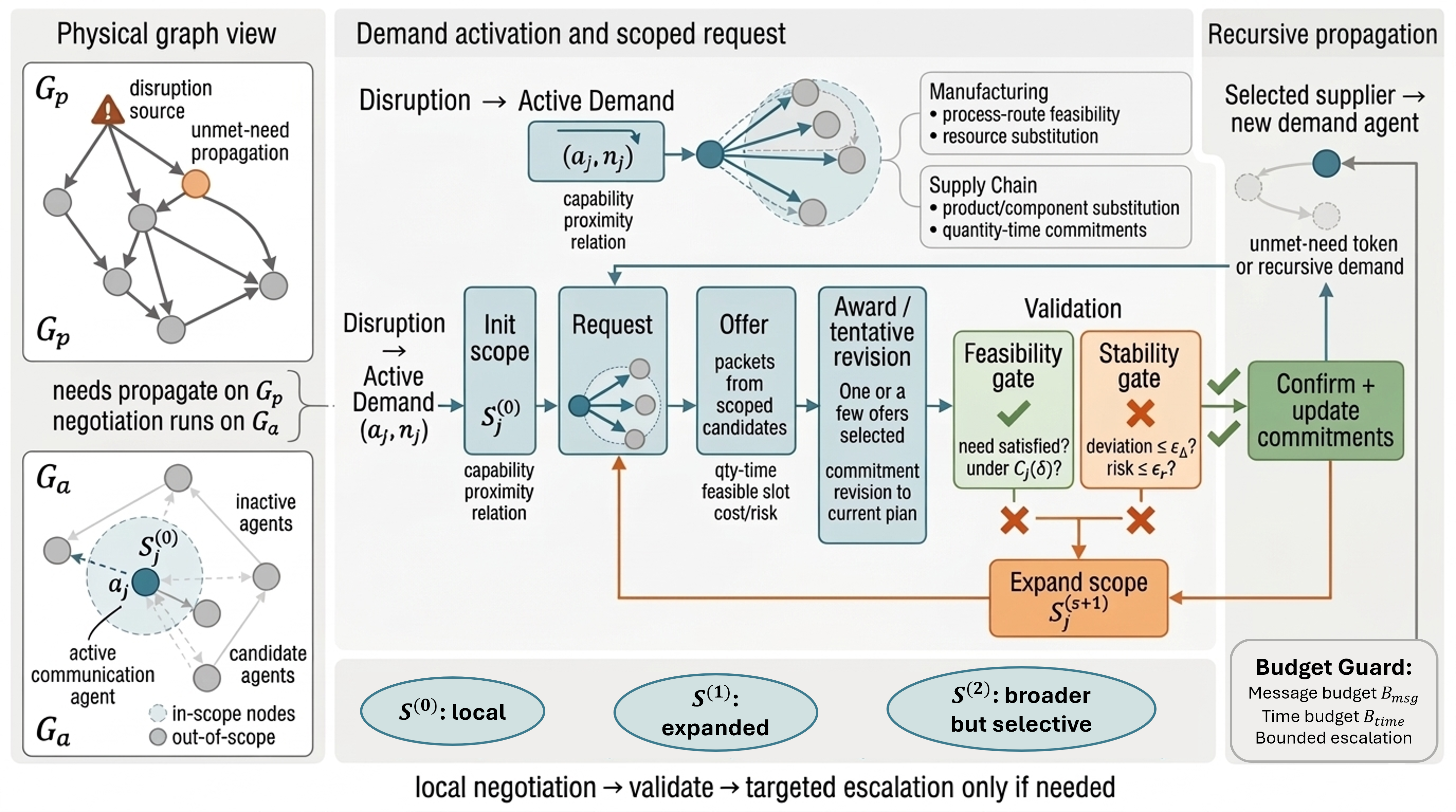}
\caption{\textbf{Gate-triggered scoped communication propagation.}
Unmet needs propagate along $G_p$ while coordination is executed on $G_a$ through contract primitives; feasibility and stability gates trigger targeted scope expansion and role transitions without global flooding.}
\label{fig:comm}
\end{figure}

This section separates \emph{what lives inside an agent} from \emph{how agents interact}.
Figure~\ref{fig:agent} specifies the unified KB/DM/CM substrate inside an individual agent.
Figure~\ref{fig:comm} specifies the \CASCADE mechanism layered on top of that substrate: scoped negotiation, validation, escalation, and propagation on the coupled graphs.
KB/DM/CM is the internal representation-and-decision blueprint; \CASCADE is the interaction mechanism that controls communication footprint under explicit budgets.
The unified substrate improves modularity and interoperability, but the paper's main mechanism contribution is the scoped interaction policy rather than the substrate alone.

\subsection{Agent architecture: KB/DM/CM}
Each agent implements three internal modules:
(i) a \textbf{Knowledge Base (KB)} for structured local beliefs, priorities, intentions, and uncertainty;
(ii) a \textbf{Decision Manager (DM)} for role-conditioned local commitment reasoning; and
(iii) a \textbf{Communication Manager (CM)} as an interface layer for physical I/O and peer messaging.
This factorization supports heterogeneous local models while preserving a common interaction interface.

\textbf{\textit{Knowledge Base.}}
Each agent $a\in A$ maintains
\begin{equation}
\label{eq:kb_def}
\mathrm{KB}_a=\{\mathcal{B}_a,\mathcal{D}_a,\mathcal{I}_a\},
\qquad
\mathcal{B}_a=\{\mathcal{X}_a,\mathcal{C}_a,\mathcal{E}_a,\mathcal{R}_a\}.
\end{equation}
$\mathcal{X}_a$ stores observable local state; $\mathcal{C}_a(\delta)$ stores local feasibility constraints and capability envelopes under disruption $\delta$; $\mathcal{E}_a$ stores the local view of $G_p$ and $G_a$; and $\mathcal{R}_a$ stores uncertainty and risk attitude.
$\mathcal{D}_a$ summarizes local priorities such as service, cost, and robustness preference.
$\mathcal{I}_a$ encodes the current local plan fragment induced by $\pi$ and its relation to $\pi_0$.

\textbf{Module Examples}: 1) each agent maintains a local capability set $\mathcal{M}_a=\{m_{a,1},\dots,m_{a,q}\},$
where each capability represents an executable local transformation, transport, or supply commitment.
In supply-chain settings, a capability can be instantiated as $(o,k)$, i.e., the ability to perform operation $o$ for product $k$.
In manufacturing settings, capabilities could be process feasibility, resource class, and local reachability, e.g., supported process types for machines and reachable cells/buffers for mobile robots.
2) The uncertainty and risk component can be represented explicitly as
$\mathcal{R}_a=\{\mathcal{U}_a,\rho_a\},$
where $\mathcal{U}_a$ denotes the local uncertainty model and $\rho_a$ denotes the local risk functional or risk attitude.
Depending on the setting, $\mathcal{U}_a$ may be represented by stochastic operation times, disruption-dependent response uncertainty, or scenario samples.
The risk functional $\rho_a$ can be instantiated as expectation, conservative worst-case evaluation, or other tail-sensitive functionals such as CVaR.

\textbf{\textit{Decision Manager.}}
The DM manages various role-conditioned local decision problems, such as request evaluation, offer generation, award selection, or local repair.
In a supplier/resource role, the DM produces feasible commitments under local constraints and uncertainty; in a demand role, it selects or mixes received offers to satisfy unmet needs while controlling local deviation and risk.

A local decision-making module may be instantiated as an optimization model, a trained policy network, or a language-guided decision module, provided that it returns commitments in the same typed format.
In the current study, the instantiated DMs are optimization-centered, with uncertainty- and risk-aware local decision models in both manufacturing and supply-chain settings.
LLM-based agent architectures provide another compatible realization of such modular decision layers.

\textbf{\textit{Communication Manager.}}
The CM routes incoming and outgoing peer messages and mediates between the DM and the physical interface.
It collects local signals, dispatches DM decisions to actuators, and standardizes peer interaction.
The question of \emph{when} and \emph{how far} communication should expand is handled by the \CASCADE mechanism below.

\subsection{\CASCADE mechanism: gate-triggered scoped propagation}

\begin{algorithm}[t]
\caption{\CASCADE: gate-triggered scoped communication for disruption replanning}
\label{alg:cascade}
\begin{algorithmic}[1]
\STATE \textbf{Input:} baseline plan $\pi_0$, disruption $\delta$, graphs $G_p,G_a$, budgets $B_{\text{msg}},B_{\text{time}}$
\STATE \textbf{Output:} revised plan $\pi$
\STATE $\pi\leftarrow\pi_0$;\quad $\mathcal{D}\leftarrow\textsc{InferDemands}(\delta,\pi_0,G_p)$
\WHILE{$\mathcal{D}\neq\emptyset$ \AND budgets not exhausted}
  \STATE Select an active demand $(a_j,n_j)\in\mathcal{D}$
  \STATE $s\leftarrow 0$;\quad initialize scope $\mathcal{S}_j^{(0)}$ via Eq.~\eqref{eq:scope_init}
  \REPEAT
    \STATE \textbf{(Request)} Send \textsc{Request}$(n_j)$ to candidates in $\mathcal{S}_j^{(s)}$; collect offers
    \STATE \textbf{(Award)} Construct a tentative revised plan $\pi^{(s)}$ from the received offers
    \STATE Evaluate $g_{\text{feas}}$ and $g_{\text{stab}}$ via Eqs.~\eqref{eq:gate_feas}--\eqref{eq:gate_stab}
    \IF{both gates pass}
      \STATE \textbf{(Confirm)} Send award/confirmation to selected suppliers; commit and set $\pi\leftarrow\pi^{(s)}$
      \STATE \textbf{(Role transition)} If commitments induce upstream needs, add new demands to $\mathcal{D}$
    \ELSE
      \STATE Expand scope via Eq.~\eqref{eq:scope_expand};\quad $s\leftarrow s+1$
    \ENDIF
  \UNTIL{both gates pass OR expansion budget exhausted}
  \STATE Remove $(a_j,n_j)$ from $\mathcal{D}$ if satisfied; otherwise terminate if budgets are exhausted
\ENDWHILE
\STATE \textbf{return} $\pi$
\end{algorithmic}
\end{algorithm}

\CASCADE is a closed-loop replanning mechanism that treats communication scope as an explicit control variable.
Mechanism execution couples the physical dependency graph $G_p$ (where unmet needs propagate) and the communication graph $G_a$ (where negotiation occurs).
A replanning episode proceeds as a sequence of scoped negotiations and commitment revisions; scope expands \emph{only} when validation gates fail.
Algorithm~\ref{alg:cascade} summarizes \CASCADE as a budgeted loop over (i) scoped negotiation on $G_a$, (ii) tentative local commitment revision, (iii) gate-based validation and escalation, and (iv) role-transition-driven propagation on $G_p$ after confirmation.

\textbf{Contract primitives.}
\CASCADE uses compact, auditable primitives:
\textsc{Request}, \textsc{Offer}, \textsc{Award}, and \textsc{Confirm}.
These primitives make message usage and negotiation rounds explicit, budgeted resources rather than hidden side effects of local search.

\textbf{Scope policy.}
Given a demand $(a_j,n_j)$, the mechanism initializes a bounded scope $\mathcal{S}_j^{(0)}\subseteq A$ on $G_a$ using capability match, structural proximity, and relation similarity:
\begin{equation}
\label{eq:scope_score}
w_{j\rightarrow b}=\alpha\,\mathrm{sim}_{\text{cap}}(n_j,\mathrm{KB}_b)+\beta\,\mathrm{prox}_{G_a}(a_j,b)+\gamma\,\mathrm{sim}_{\text{rel}}(a_j,b),
\end{equation}
\begin{equation}
\label{eq:scope_init}
\mathcal{S}_j^{(0)}
=
\{b:\mathrm{dist}_{G_a}(a_j,b)\le h_0\}\ \cap\ \mathrm{TopK}(w_{j\rightarrow \cdot},K_0).
\end{equation}
If escalation is needed, scope expands according to a schedule $(h_s,K_s)_{s\ge 0}$:
\begin{equation}
\label{eq:scope_expand}
\mathcal{S}_j^{(s+1)}=\mathcal{S}_j^{(s)}\cup
\Big(\{b:\mathrm{dist}_{G_a}(a_j,b)\le h_{s+1}\}\cap \mathrm{TopK}(w_{j\rightarrow \cdot},K_{s+1})\Big).
\end{equation}
The main text keeps this formulation domain-agnostic; concrete environment-specific instantiations are deferred to the appendix.
In manufacturing, capability match corresponds to process-route feasibility and local resource substitutability, while proximity reflects resource adjacency and transport reachability in the production system.
In the supply chain, capability match corresponds to feasible product/component substitution and commitment compatibility, while proximity reflects dependency closeness on the production--logistics network.
Relation similarity in both settings captures whether two agents occupy structurally or historically compatible coordination roles, allowing the same score to prioritize plausible local substitutes without collapsing immediately to broad broadcast.

\textbf{Validation gates.}
Let $\pi^{(s)}$ denote the tentative plan after negotiation within scope $\mathcal{S}_j^{(s)}$.
Two gates regulate acceptance and escalation:
\begin{equation}
\label{eq:gate_feas}
g_{\text{feas}}(a_j,n_j,\pi^{(s)},\delta)=\mathbf{1}\!\left[\exists\,u_j \ \text{s.t.}\ (u_j,\pi^{(s)})\in \mathcal{C}_j(\delta)\ \wedge\ \text{$n_j$ is satisfied}\right],
\end{equation}
\begin{equation}
\label{eq:gate_stab}
g_{\text{stab}}(a_j,\pi^{(s)},\pi_0)=\mathbf{1}\!\left[d_j(\pi^{(s)},\pi_0)\le \varepsilon_\Delta\ \wedge\ \rho_j(\pi^{(s)})\le \varepsilon_r\right].
\end{equation}
The feasibility gate checks whether the current scope can actually satisfy the induced need under disruption-aware constraints.
The stability gate checks whether the induced revision remains within acceptable deviation and risk.
Confirmation is final after both gates pass; before that, offers are treated as tentative and may be discarded when scope expands via Eq.~\eqref{eq:scope_expand}.

\textbf{Role transition and propagation.}
If an awarded offer induces upstream needs (e.g., additional materials, transport, or substitute capacity), the selected supplier becomes a new demand agent.
This role transition drives propagation on $G_p$, while communication on $G_a$ remains explicitly scoped.
The mechanism, therefore, implements a bounded loop:
local negotiation $\rightarrow$ validation $\rightarrow$ targeted escalation only if needed.

\section{Experiments}
This section evaluates \CASCADE in two industrial domains:
(i) manufacturing disruption replanning (rescheduling and resource reallocation), and
(ii) supply-chain disruption response (flow and commitment revision).
Across both domains, agents operate on a physical dependency graph $G_p$ where unmet needs propagate and exchange messages on a communication graph $G_a$ where negotiation is executed.
A disruption $\delta$ changes local capabilities $\mathcal{C}_a(\delta)$ and may induce unmet needs that propagate through dependencies.

\subsection{Environments and disruption regimes}
\textbf{Manufacturing.}
We study a tightly coupled production environment with $20$ machine agents and $6$ mobile-robot agents~\citep{bi2024dynamic}.
Disruptions such as machine breakdowns and availability changes invalidate a baseline production schedule, forcing coordinated operation reassignment under feasibility constraints.
The manufacturing setting additionally includes uncertainty and a risk-aware variant used for robustness analysis.

\textbf{Supply chain.}
We study an interdependent automotive-cockpit production--logistics network~\citep{bi2023distributed, bi2024heterogeneous}.
Two disruption regimes are reported on this shared network.
The first is a network-structured supplier-loss setting, used to analyze how structural attributes affect propagation footprint and communication.
The second is a stochastic lead-time disruption setting with heterogeneous risk attitudes, used to analyze robustness under uncertainty.
Exact system specifications, disruption protocols, and graph definitions are deferred to the appendix.

\begin{table}[t]
\centering
\footnotesize
\setlength{\tabcolsep}{3pt}
\caption{Compact summary of the two experimental settings used in the main text. Detailed environment definitions, disruption-generation rules, and metric formulas are provided in the appendix.}
\label{tab:exp_settings}
\begin{tabularx}{\columnwidth}{
>{\raggedright\arraybackslash}p{1.75cm}
>{\raggedright\arraybackslash}X
>{\raggedright\arraybackslash}p{1.75cm}
>{\raggedright\arraybackslash}X}
\toprule
\textbf{Setting} & \textbf{System and disruption regime} & \textbf{Coordination} & \textbf{Main reported signals / protocol} \\
\midrule
Manufacturing
& Modified Intel Mini-Fab; two product types with fixed process routes; disruptions are machine breakdowns with operation-time uncertainty.
& Centralized vs.\ \CASCADE
& Cycle time, communication $M$, runtime $T$; incident and risk statistics in the disruption-aware case; 5 trials. \\
\midrule
Supply chain
& Automotive cockpit supply-chain network. Two disruption regimes: (i) single supplier-loss on the nominal flow plan; (ii) lead-time disruption with different added time attributes.
& Centralized vs.\ \CASCADE; neutral vs.\ averse
& Overage cost $O$, communication $M$, and footprint proxies $(N_c,N_a)$; delayed-product distribution and cost/lateness $(C_{\mathrm{dm}},L_{\mathrm{dm}},J_{\mathrm{dm}})$. \\
\bottomrule
\end{tabularx}
\end{table}

\subsection{Experimental protocol and metrics}
We compare \textbf{Centralized} replanning (global coordination) against \textbf{Distributed} replanning implemented by \CASCADE.
Table~\ref{tab:exp_settings} summarizes the evaluation regimes and the scale at which each diagnostic is constructed.


\textbf{Budgets and accounting.}
Communication is measured by the number of point-to-point message exchanges $M$.
Each \textsc{Request}, \textsc{Offer}, \textsc{Award}, and \textsc{Confirm} sent to one recipient counts as one exchange; multicast messages are counted per recipient.
Centralized coordination is accounted for separately using a domain-specific aggregation of information-collection and update-notification steps, whereas distributed coordination counts only the explicitly executed negotiation messages needed to produce the revised plan.
Latency $T$ is measured as replanning runtime under the same implementation setting.
All communication results should therefore be read as message-exchange accounting rather than bit-level or semantic-content accounting.
The detailed domain-specific protocol, scenario counts, and accounting conventions are provided in the appendix.

\textbf{Implementation contingency.}
In the manufacturing implementation, when the local candidate set is empty, the environment permits escalation to a central controller or human manager as an external fallback.
We report this behavior explicitly in the appendix; it is an implementation contingency rather than the intended scoped interaction path and is not treated as a mechanism contribution.

\textbf{Quality and feasibility.}
In manufacturing, the primary quality proxy is disruption-recovery quality measured through cycle-time behavior together with incident and risk statistics.
In supply chain, quality is measured by overage cost $O$ and feasibility status, with additional lateness and cost--delay metrics in the stochastic setting.
Formal definitions are collected in the appendix.

\subsection{Unified diagnostics}
We report three complementary diagnostics that expose mechanism-level trade-offs rather than only task-level averages.
The three views are deliberately chosen to answer different questions raised by budgeted industrial coordination.
R1 asks whether better recovery quality can be obtained without paying the full latency and communication cost of broad coordination.
R2 asks whether the induced communication footprint follows interpretable structural properties of the disrupted network rather than expanding arbitrarily.
R3 asks whether risk-aware acceptance and stabilization policies improve tail behavior under stochastic disruption, rather than merely shifting average nominal cost.

\begin{figure}[t]
\centering
\includegraphics[width=0.9\linewidth]{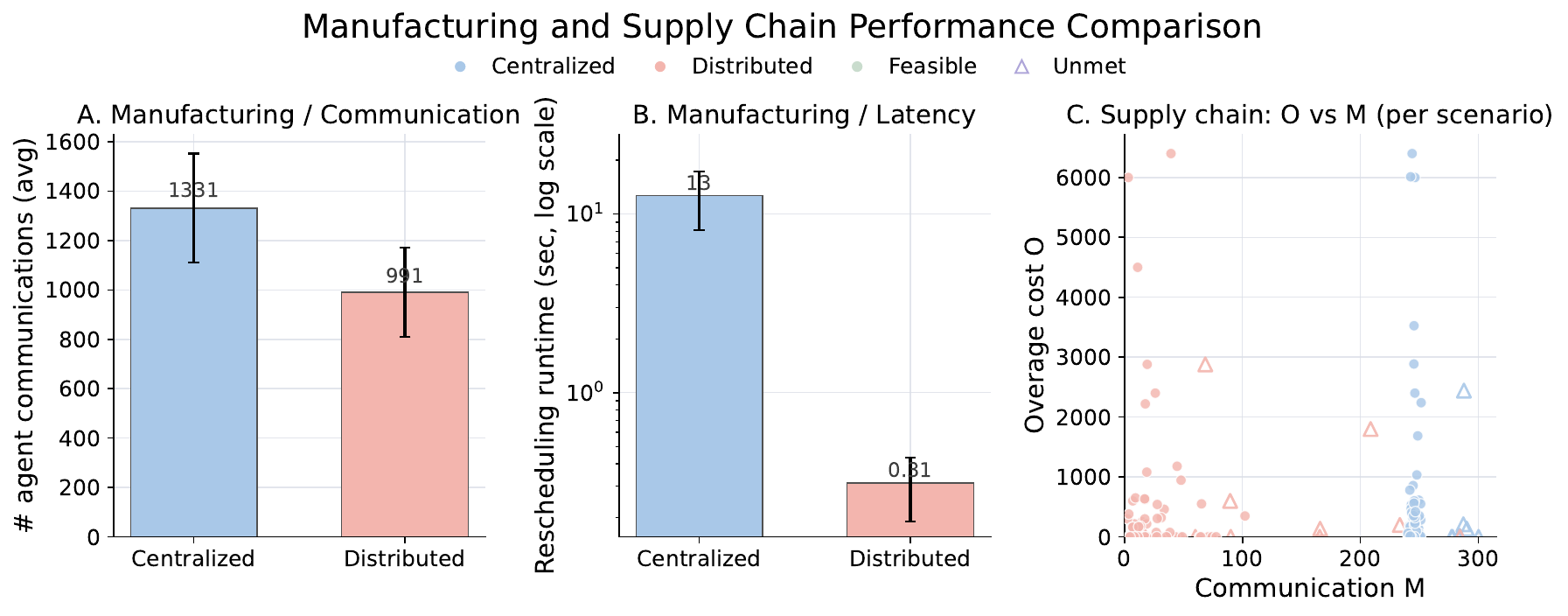}
\caption{R1: Quality--latency--communication frontier. The axes are domain-specific (cycle-time-based recovery proxy in manufacturing; overage cost with feasibility labels in supply chain), but the shared question is how much recovery quality is achieved per unit communication and latency.}
\label{fig:frontier}
\end{figure}

\textbf{R1: Quality--latency--communication frontier.}
Figure~\ref{fig:frontier} summarizes the relationship between task quality, communication $M$, and latency $T$.
In manufacturing, the quality axis is a cycle-time-based recovery proxy; in supply chain, it is scenario-level overage cost $O$ together with feasibility labels.
The purpose of R1 is not to force identical units across domains, but to ask the same mechanism question in both settings: how much recovery quality is achieved per unit communication and latency.
In manufacturing, \CASCADE reduces replanning latency substantially while requiring fewer communications than centralized coordination.
In the supply chain, the $O$--$M$ scatter exposes scenario-level trade-offs and feasible versus unmet cases. 

\begin{figure}[t]
\centering
\includegraphics[width=\linewidth]{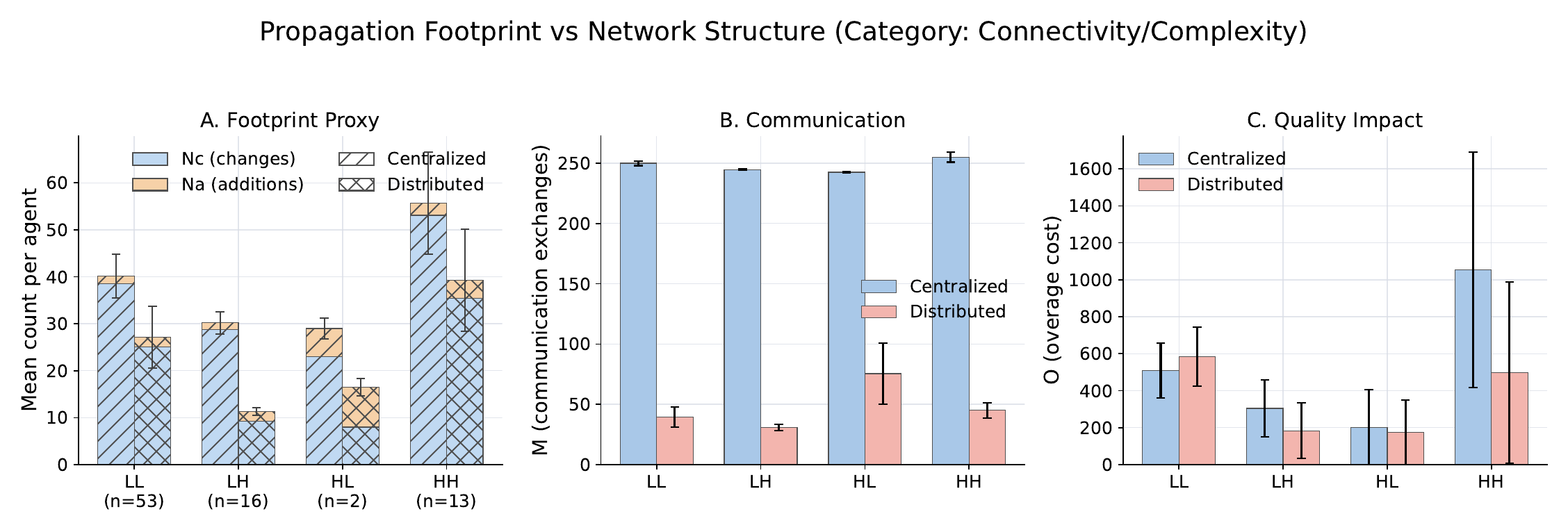}
\caption{R2: Propagation footprint vs.\ network structure. Footprint proxies ($N_c$, $N_a$), communication $M$, and quality impact $O$ are stratified by network-attribute groups (LL/LH/HL/HH). The grouped analysis is restricted to the 73 supplier-loss scenarios in which both centralized and distributed recovery are feasible; group sizes are LL$=53$, LH$=16$, HL$=2$, and HH$=13$.}
\label{fig:attributes}
\end{figure}

\textbf{R2: Propagation footprint and network structure (supply chain).}
Figure~\ref{fig:attributes} links propagation footprint to structural attributes in the environment model.
We instantiate the footprint using proxy counts such as commitment changes $N_c$, added commitments $N_a$, communication $M$, and quality impact $O$.
These proxies are appropriate because they capture how far a local disruption response must spread in order to restore feasibility: more changed commitments, more activated entities, and more messages all indicate deeper cascade depth.
The four groups (LL/LH/HL/HH) correspond to a low/high stratification of connectivity and redundancy.
This diagnostic explains \emph{when} deeper cascades become necessary: tightly coupled or low-redundancy regions induce larger footprint and communication, while redundant neighborhoods admit earlier termination.

\begin{figure}[t]
\centering
\includegraphics[width=\linewidth]{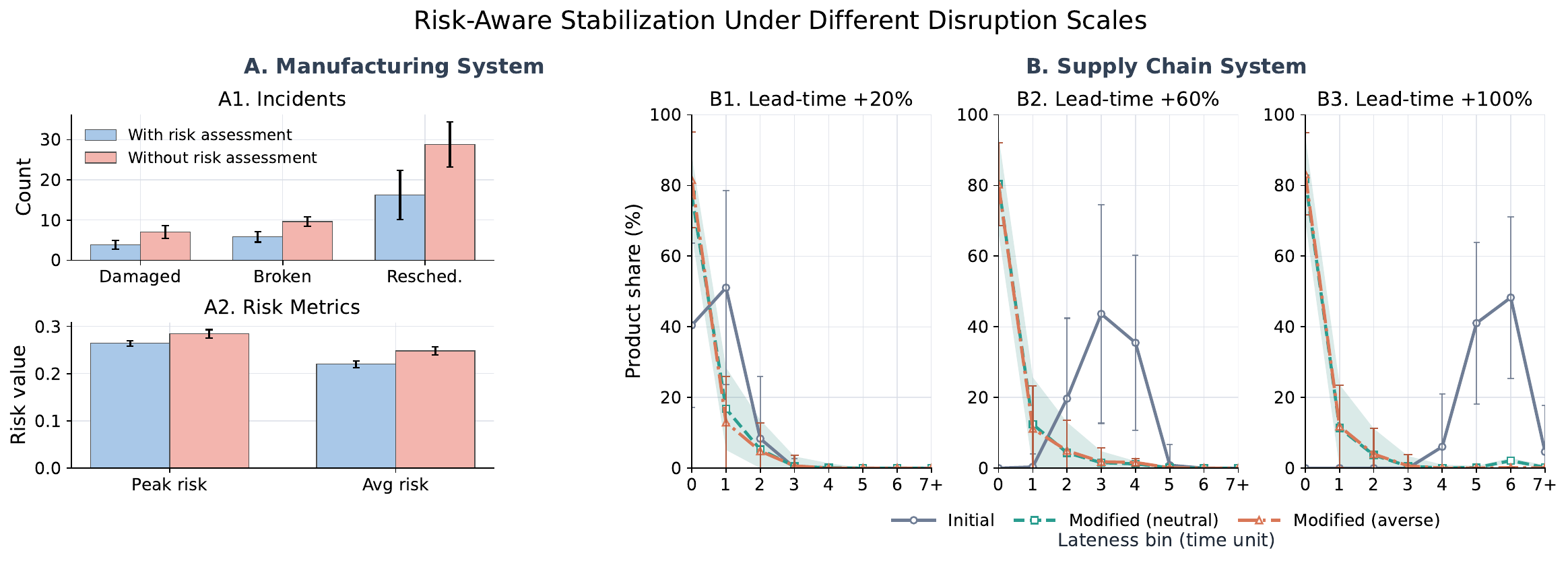}
\caption{R3: Robustness under uncertainty. The manufacturing panel summarizes five trials, while the supply-chain panel reports the fixed disrupted-subnetwork study. Across both settings, risk-aware stabilization improves tail behavior and reduces brittle recovery under stochastic disruptions.}
\label{fig:risk}
\end{figure}

\begin{table}[t]
\centering
\footnotesize
\setlength{\tabcolsep}{8pt}
\caption{R3 robustness summary across the two industrial domains. Panel A reports the effect of risk assessment in manufacturing recovery (five trials). Panel B reports the effect of risk attitude in the supply-chain lead-time disruption setting. Negative $\Delta$ indicates improvement for loss/risk metrics; negative $\Delta J$ indicates that the Averse policy improves the objective relative to Neutral.}
\label{tab:r3_unified}
\begin{tabular}{lcccc}
\toprule
\multicolumn{5}{c}{\textbf{Panel A: Manufacturing robustness}} \\
\midrule
\textbf{Metric} & \textbf{With risk} & \textbf{Without} & \textbf{$\Delta$} & \textbf{$\Delta$\%} \\
\midrule
\# damaged products
& 3.80$\pm$1.10 & 7.00$\pm$2.12 & -3.20 & -45.7\% \\
\# broken machines
& 4.52$\pm$1.23 & 7.84$\pm$2.49 & -3.32 & -42.3\% \\
\# rescheduled processes
& 4.05$\pm$1.19 & 7.20$\pm$2.47 & -3.15 & -43.8\% \\
Peak risk value
& 0.02$\pm$0.04 & 0.03$\pm$0.04 & -0.01 & -33.3\% \\
Average risk value
& 0.03$\pm$0.02 & 0.06$\pm$0.04 & -0.03 & -50.0\% \\
\midrule
\multicolumn{5}{c}{\textbf{Panel B: Supply-chain robustness}} \\
\midrule
\textbf{Rate} & \textbf{Risk} & \textbf{$C_{\mathrm{dm}}$} & \textbf{$L_{\mathrm{dm}}$} & \textbf{$J_{\mathrm{dm}}$ / $\Delta J$} \\
\midrule
Lead time +0\% 
& Neutral & 31,120 & 0 & 31,120 / -- \\
\midrule
Lead time +20\%
& Neutral & 28,386 & 1 & 128,386 / -- \\
& Averse  & 28,416 & 1 & 128,416 / +30 (+0.02\%) \\
\midrule
Lead time +60\%
& Neutral & 28,386 & 6 & 628,386 / -- \\
& Averse  & 29,342 & 5 & 529,342 / -99,044 (-15.76\%) \\
\midrule
Lead time +100\%
& Neutral & 28,792 & 8 & 828,792 / -- \\
& Averse  & 29,208 & 3 & 329,208 / -499,584 (-60.28\%) \\
\bottomrule
\end{tabular}
\end{table}

\textbf{R3: Robustness under uncertainty via risk-aware stabilization.}
Figure~\ref{fig:risk} summarizes robustness under stochastic disruptions.
In manufacturing, robustness is reflected by incident counts and schedule-risk summaries with and without risk assessment.
In the supply chain, robustness is reflected by delayed-product distributions and cost--lateness trade-offs under increasing disruption intensity.
The common mechanism interpretation is that risk-aware stabilization changes the acceptance policy for local commitments, improving tail behavior even when the nominal cost may increase slightly.

\section{Discussion}
\textbf{Why scoped cascades fit industrial replanning.}
Industrial replanning often crosses organizational boundaries and must obey strict latency budgets.
In such settings, communication is not a free software detail; it is an operational resource.
By exposing scope schedules and gate thresholds as explicit knobs, \CASCADE\ supports auditable escalation under practical constraints such as privacy, supplier contracts, and bounded negotiation windows.

\textbf{Where the mechanism matters most.}
The diagnostics suggest that scoped cascades are most valuable in the intermediate regime where disruptions can propagate beyond immediate neighbors but do not require full-network mobilization in every case.
In tightly coupled or low-redundancy regions, local fixes often fail, and deeper propagation becomes necessary.
In more redundant neighborhoods, the same mechanism can terminate early and avoid unnecessary communication.
Conversely, when the feasible substitute set is already contained in a very small neighborhood, fixed local coordination may already suffice; when almost the entire network must be mobilized, the mechanism approaches broad escalation.
This boundary is important: \CASCADE\ matters most when \emph{adaptive} scope control is the deciding factor.

\textbf{How the current evidence should be read.}
The present results support a mechanism-design claim rather than a complete algorithmic ranking.
The diagnostics are intended to isolate the value of explicit scope control and gate-triggered escalation across two industrial domains, not yet to substitute for a full benchmark study against all distributed coordination baselines.
Fixed-scope distributed negotiation and ungated schedule-based expansion are the most important missing comparators for testing whether adaptive gated scope is genuinely necessary, while stronger DCOP-style baselines would help separate scoped control from decentralized replanning more broadly.
Likewise, the current study exposes scope schedules and gate thresholds as auditable knobs, but does not yet provide a systematic sensitivity sweep across $(h_s,K_s)$ and $(\varepsilon_\Delta,\varepsilon_r)$ to test deployment stability.

\textbf{Remaining limitations.}
First, the evaluation spans two representative industrial domains, but it is not yet a fully standardized public benchmark suite.
Second, communication is accounted for at the level of message exchanges rather than bit-level or content-level cost; integrating task-level compression and richer network constraints remains an important extension.
Third, the stochastic supply-chain analysis focuses on a single disrupted supplier and a limited heterogeneous-risk setting; broader topological regimes and richer combinations of conflicting agent risk attitudes remain open.

\section{Conclusion}
\CASCADE is a budgeted industrial replanning mechanism that makes the communication footprint explicit and controllable.
By combining a unified KB/DM/CM agent substrate with contract-style negotiation and gate-triggered scope expansion, it enables stepwise, auditable coordination without immediate global flooding.
Across manufacturing rescheduling and supply-chain disruption response, the unified diagnostics provide mechanism-level evidence for a common quality--latency--communication frontier, relate propagation footprint to network structure, and show that risk-aware stabilization can improve robustness under stochastic disruptions.

The present results should be read as support for a design principle rather than a complete benchmark verdict.
The main lesson is that communication scope should not be treated as fixed or free in industrial disruption response.
It should be made explicit, validated before confirmation, and escalated only when the current coordination neighborhood is insufficient.

\bibliography{iclr2026_conference_camera_ready}
\bibliographystyle{iclr2026_conference}

\appendix

\section{Full Mechanism Specification}
\label{app:mechanism}

\subsection{Unified agent blueprint}
Each agent implements three internal modules: a \emph{Knowledge Base} (KB), a \emph{Decision Manager} (DM), and a \emph{Communication Manager} (CM). The KB stores structured local beliefs, desires, and intentions,
\begin{equation}
\mathrm{KB}_a = \{B_a, D_a, I_a\}, \qquad B_a = \{X_a, C_a, E_a, R_a\},
\end{equation}
where $X_a$ denotes observable local state, $C_a(\delta)$ denotes local feasibility constraints and capability envelopes under disruption $\delta$, $E_a$ denotes the local view of the physical dependency graph and the communication graph, and $R_a$ stores uncertainty and risk attitude. Desires $D_a$ summarize local priorities such as service level, cost sensitivity, and robustness preference, while intentions $I_a$ encode the current local plan fragment induced by the baseline plan $\pi_0$ and the revised plan $\pi$.

The DM solves role-conditioned local optimization problems. When an agent acts as a supplier/resource, it generates offers by solving a constrained local problem over feasible commitments; when it acts as a demand agent, it selects offers to satisfy unmet needs while controlling deviation and risk. The CM provides the interface to the physical layer and to peer agents. It collects local signals, dispatches DM decisions to actuators, and routes incoming and outgoing peer messages.

\subsection{Offer generation and award selection}
Given an incoming request, a supplier-side agent solves a local risk-aware commitment problem of the form
\begin{equation}
\min_{z \in Z_i} \; \rho_{\omega \sim R_i}\big[J_i(z;\omega)\big]
\qquad \text{s.t.} \quad z \in C_i(\delta),
\label{eq:app_offer}
\end{equation}
where $z$ parameterizes the proposed commitment (for example, quantity--time commitments and auxiliary local actions), and $\rho$ is a risk functional such as expectation or CVaR. The resulting \textsc{offer} includes feasibility and robustness summaries.

After collecting offers $O_j^{(s)}$ from the current scope $S_j^{(s)}$, the demand-side agent solves an award problem
\begin{equation}
\min_y \, \sum_{b \in S_j^{(s)}} y_{jb} J_{jb}(o_{jb})
+ \lambda_{\Delta} d_j\!\left(\pi^{(s)}(y),\pi_0\right)
+ \lambda_r \rho_j\!\left(\pi^{(s)}(y)\right)
\label{eq:app_award_obj}
\end{equation}
subject to
\begin{equation}
\mathrm{NEEDSAT}\!\left(n_j,y,O_j^{(s)}\right)=1,
\qquad
\mathrm{LOCALFEAS}\!\left(y;C_j(\delta)\right)=1,
\label{eq:app_award_con}
\end{equation}
with $y_{jb} \in \{0,1\}$ for single-award selection or $y_{jb}\in[0,1]$ for splittable commitments. Here $J_{jb}(o_{jb})$ is the local evaluated utility/cost of selecting offer $o_{jb}$, $d_j(\cdot,\cdot)$ is local plan deviation relative to $\pi_0$, and $\rho_j(\cdot)$ is a robustness/risk score.

\subsection{Scoped communication policy}
CASCADE treats communication scope as an explicit control variable on the communication graph $G_a=(A,L)$. For a demand $(a_j,n_j)$, the initial scope is selected using a weighted score
\begin{equation}
w_{j \rightarrow b} = \alpha\,\mathrm{sim}_{\mathrm{cap}}(n_j,\mathrm{KB}_b)
+ \beta\,\mathrm{prox}_{G_a}(a_j,b)
+ \gamma\,\mathrm{sim}_{\mathrm{rel}}(a_j,b),
\label{eq:app_scope_score}
\end{equation}
where the three terms respectively capture capability match, structural proximity, and relationship similarity. The initial scope is
\begin{equation}
S_j^{(0)} = \{b : \mathrm{dist}_{G_a}(a_j,b) \le h_0\}
\cap \mathrm{TopK}(w_{j \rightarrow \cdot},K_0).
\label{eq:app_scope_init}
\end{equation}
If escalation is needed, the scope expands according to a schedule $(h_s,K_s)_{s\ge 0}$:
\begin{equation}
S_j^{(s+1)} = S_j^{(s)} \cup
\Big(\{b : \mathrm{dist}_{G_a}(a_j,b) \le h_{s+1}\}
\cap \mathrm{TopK}(w_{j \rightarrow \cdot},K_{s+1})\Big).
\label{eq:app_scope_expand}
\end{equation}
Operational constraints such as organizational boundaries, privacy restrictions, or supplier-policy masks can be enforced by removing candidate agents or communication edges from $G_a$ before ranking.

\subsection{Validation gates, propagation, and stopping}
For the tentative plan $\pi^{(s)}$ generated from the current scope $S_j^{(s)}$, CASCADE evaluates two gates:
\begin{equation}
 g_{\mathrm{feas}}\!\left(a_j,n_j,\pi^{(s)},\delta\right)=1
 \iff
 \exists\,u_j \;\text{s.t.}\; (u_j,\pi^{(s)})\in C_j(\delta)
 \;\wedge\; n_j\;\text{is satisfied},
\label{eq:app_gfeas}
\end{equation}
\begin{equation}
 g_{\mathrm{stab}}\!\left(a_j,\pi^{(s)},\pi_0\right)=1
 \iff
 d_j\!\left(\pi^{(s)},\pi_0\right)\le \varepsilon_{\Delta}
 \;\wedge\;
 \rho_j\!\left(\pi^{(s)}\right)\le \varepsilon_r.
\label{eq:app_gstab}
\end{equation}
If either gate fails, the mechanism expands scope via Eq.~\eqref{eq:app_scope_expand}. If an awarded offer induces upstream needs (e.g., additional materials, transport, or substitute capacity), the selected supplier becomes a new demand agent. In this way, unmet needs propagate on the physical dependency graph $G_p$, while negotiation remains explicitly scoped on $G_a$.

A replanning episode terminates when all active demands are satisfied, when the expansion budget is exhausted, or when hard message/time budgets are exhausted. In the current paper, the operational budgets are the message budget $B_{\mathrm{msg}}$ and wall-clock deadline $B_{\mathrm{time}}$.

\section{Manufacturing Environment and Disruption Model}
\label{app:manufacturing}

\subsection{Facility layout and agentization}
The manufacturing environment is a modified Intel Mini-Fab instance implemented in Repast Symphony. The simulated facility contains two infinite-capacity buffers (\texttt{Entry} and \texttt{Exit}), $20$ machine agents, and $6$ mobile-robot agents. Machines support six process types, P1--P6. Operation costs are measured in RepastS ticks and range from $110$ to $200$ depending on the process and resource.

Each machine agent is associated with a subset of process capabilities and a workspace class (small or large). Each robot agent is associated with a reachability set over cells and buffers. In the unified AIMS view, this environment is treated as a tightly coupled production graph with $26$ resource agents.

\subsection{Products and nominal schedule}
Two product types are used:
\begin{itemize}
    \item \textbf{S-product:} $\mathrm{P1} \rightarrow \mathrm{P2} \rightarrow \mathrm{P3} \rightarrow \mathrm{P6}$.
    \item \textbf{L-product:} $\mathrm{P1} \rightarrow \mathrm{P3} \rightarrow \mathrm{P4} \rightarrow \mathrm{P5}$.
\end{itemize}
Machines labeled \texttt{L} can process both L-products and S-products, while machines labeled \texttt{S} process only S-products. Products enter from \texttt{Entry} and leave through \texttt{Exit} after completing their required routes.

The nominal workload contains $50$ L-products and $50$ S-products, released alternately every $30$ ticks starting at tick $10$. The initial production schedule is generated at approximately $50\%$ resource utilization so that local rescheduling has available slack under disruptions.

\subsection{Disruption regime}
The manufacturing disruption regime models machine breakdowns together with uncertainty in operation time. Machine breakdown probabilities at the start of simulation range from $3.3\%$ to $10\%$. When a machine breaks down, a rescheduling process is triggered. The mean time to repair ranges from $1000$ to $1500$ ticks. If a breakdown occurs while a machine is processing a product, that product is damaged and cannot be recovered.

\subsection{Rescheduling semantics}
The local manufacturing replanning problem replaces an affected event sequence $s_d$ by a new event sequence $s_{\mathrm{new}}$ such that the relevant product transition is preserved:
\begin{equation}
\mathrm{Tr}(x_{\mathrm{prior}}, s_{\mathrm{new}})=x_{\mathrm{post}}.
\end{equation}
Only the shortest affected sequence is replaced, which keeps replanning local and preserves as much of the nominal schedule as possible. Transportation events may need to be inserted when a transformation operation is moved to a different machine.

Given a candidate resource schedule, the earliest feasible start time heuristic $H(I,t,\delta,e)$ inserts a new event $e$ into a resource idle-interval set $I$ while allowing at most one local shift of an already scheduled operation. This heuristic is used to keep rescheduling local and to reduce induced deviation from the original schedule.

\subsection{Manufacturing risk model}
The risk-aware manufacturing setting evaluates two risks for a candidate replacement sequence:
\begin{itemize}
    \item \textbf{Operational-delay risk} $R_1$: the likelihood that inserting a new event causes downstream delays for other products already assigned to the selected resource.
    \item \textbf{Breakdown risk} $R_2$: the likelihood that assigning the new event increases the probability of a future resource breakdown.
\end{itemize}
The total risk is represented by a weighted sum,
\begin{equation}
R(s_{\mathrm{new}})=w_1 R_1(s_{\mathrm{new}})+w_2 R_2(s_{\mathrm{new}}),
\end{equation}
with the reported case using $w_1=0.2$ and $w_2=0.8$. This weighting emphasizes avoidance of brittle recovery sequences that push machines close to failure.

\section{Supply-Chain Environment and Disruption Models}
\label{app:supplychain}

\subsection{Cockpit supply-chain network}
The supply-chain testbed is an automotive cockpit network. Vehicle cockpits are the final products and are assembled from components, parts, and raw materials. The network contains three vehicle models, and each model requires one, two, or three cockpit styles. Each auto assembly plant produces a single vehicle model, while each cockpit assembly plant can produce multiple cockpit styles.

Each cockpit requires ten components. For cockpits of the same vehicle model, the cluster, substrate, glove box, HVAC system, cross-car beam, and steering column are shared. Each cockpit style has a style-specific infotainment system, wiring harness, and bezel combination, while all cockpits share a common airbag type. Upstream component production depends on parts and raw materials, and different components may share part/material suppliers.

The full network contains $117$ supplier/customer agents and $413$ transportation agents. The supplier/customer layer contains $5$ customers, $3$ cockpit assembly plants, $31$ component suppliers, $62$ part suppliers, and $16$ raw-material suppliers. Each supplier or transportation agent has its own cost and capacity attributes.

\subsection{Graph model and agent roles}
The supply chain is modeled as a physical flow graph $G=(V,E)$ whose vertices are enterprise entities and whose edges are transportation/flow channels. In the agentized view, both vertices and edges can be represented as agents in the communication network. Roles include customer, distributor, OEM, tier supplier, and transporter. In the current AIMS exposition, the mechanism is instantiated over the active disruption-response subnetwork induced by the scenario under study, while keeping the same cockpit-network semantics.

\subsection{Network-structured disruption setting}
The network-structured setting studies a single supplier-loss disruption. The initial flow plan is computed by a centralized optimization model and then perturbed by removing one supplier agent that is active in the nominal plan. The scenario-generation rules are:
\begin{itemize}
    \item the lost agent must have production tasks in the initial plan;
    \item exactly one supplier agent is disrupted in each scenario;
    \item production and transportation can exceed nominal capacity by up to $30\%$ at an additional $50\%$ unit cost.
\end{itemize}
In the nominal plan, $84$ agents carry production tasks, yielding $84$ disruption scenarios. Of these, $73$ scenarios admit feasible recovery plans for both centralized and distributed coordination and are retained for the attribute-stratified analysis.

\subsection{Lead-time disruption and heterogeneous-risk setting}
The stochastic supply-chain setting uses the same cockpit network with additional temporal attributes. Lead times are attached to suppliers and cockpit assemblers, and delivery deadlines are attached to customer demands. The network is initialized with an optimal baseline flow plan.

The disrupted supplier is \texttt{cluster\_sup\_3} (denoted $S3$), which supplies three cluster products,
\begin{equation}
K_d = \{\texttt{cluster\_1},\texttt{cluster\_2},\texttt{cluster\_3}\}.
\end{equation}
Its downstream demand agents are the three cockpit suppliers
\begin{equation}
A_{\mathrm{dm}} = \{A_1,A_2,A_3\},
\end{equation}
and alternative cluster suppliers are denoted by $S1$, $S2$, and $S4$. The disruption increases the lead time of $S3$ by $20\%$, $60\%$, or $100\%$.

\subsection{Uncertainty model and distributed decision making}
The stochastic lead-time case uses a discrete-event simulation. Simulation starts from upstream agents with outflows at time $0$. For each outflow, lead time is sampled from a known normal distribution. A downstream agent begins production only after all required components have arrived, so its completion time is determined by the latest required arrival among its incoming components. The process is iterated from upstream to downstream until final cockpits are delivered to all customers. Out-of-sample evaluation uses $300$ independent replications.

Risk heterogeneity enters at both offer generation and supplier selection. A supplier agent solves a local stochastic response problem under uncertainty in production capacity and lead time. A demand agent solves a supplier-selection problem using received responses, while modeling trust in a supplier response as uncertainty around the offered quantity and arrival time. In the case study, risk-neutral decision making optimizes the expected objective via sample-average approximation, whereas risk-averse decision making optimizes a conservative worst-case variant.

\section{Metric Definitions}
\label{app:metrics}

\subsection{Plan deviation and communication}
At the mechanism level, the revised plan is scored by a generic objective
\begin{equation}
J(\pi) = J_{\mathrm{service}}(\pi) + \lambda_c J_{\mathrm{cost}}(\pi)
+ \lambda_{\Delta} d(\pi,\pi_0) + \lambda_{\mathrm{msg}} M(\pi),
\end{equation}
where $d(\pi,\pi_0)$ is plan deviation and $M(\pi)$ is communication on $G_a$. A common instantiation of plan deviation is the sum of changed commitments plus local edit distances:
\begin{equation}
 d(\pi,\pi_0) = \sum_{(i\rightarrow j,k)\in E_p} \mathbf{1}
 \big[(c_{ij}^k,\tau_{ij}^k) \neq (c_{ij}^{k,0},\tau_{ij}^{k,0})\big]
 + \sum_{a\in A} d_a(u_a,u_a^0).
\end{equation}
Communication is the total number of exchanged messages during replanning,
\begin{equation}
M(\pi)=\sum_{m\in \mathcal{M}(\pi)} 1.
\end{equation}

\subsection{Manufacturing metrics}
The manufacturing studies use the following metrics:
\begin{itemize}
    \item \textbf{Cycle time:} completion time per product; used as the main recovery-quality proxy in the manufacturing frontier.
    \item \textbf{Communication $M$:} total request/response/award/confirm style exchanges during the rescheduling process.
    \item \textbf{Latency $T$:} total wall-clock runtime of the replanning implementation in seconds.
    \item \textbf{Damaged products:} number of products damaged due to breakdown during processing.
    \item \textbf{Broken machines:} number of machine breakdown events.
    \item \textbf{Rescheduled processes:} number of triggered rescheduling episodes.
    \item \textbf{Peak and average risk values:} summary statistics of the selected replacement schedules.
\end{itemize}
In the unified diagnostics, cycle time is used as the main quality proxy for R1, while damaged products, broken machines, rescheduled processes, and risk values support R3.

\subsection{Supply-chain structural attributes}
For the supply-chain environment, four structural attributes characterize the role of a disrupted agent:
\paragraph{Connectivity.}
\begin{equation}
C_i = \sum_{a_j\in V} b_{ij} + b_{ji},
\end{equation}
where $b_{ij}=1$ if edge $(i,j)$ carries material/product flow and $0$ otherwise.

\paragraph{Depth.}
\begin{equation}
D_i = \max_{a_j\in \mathrm{Customer}} d(a_i,a_j),
\end{equation}
where $d(\cdot,\cdot)$ is geodesic distance.

\paragraph{Capability redundancy.}
For capability $m$,
\begin{equation}
R_i(m)=\big|\{a_j\mid a_j \text{ has capability } m,\; a_j\in A\setminus\{a_i\}\}\big|.
\end{equation}

\paragraph{Production complexity.}
\begin{equation}
P_i = \left|\bigcup_{k\in K_i}\{k_f \mid k\in Z(k_f),\; \forall k_f\in K_f\}\right|
+ \left|\bigcup_{k\in K_i} Z(k)\right|,
\end{equation}
where $K_i$ is the product set of agent $a_i$, $K_f$ is the final product set, and $Z(k)$ maps a product to the required components/materials.

\subsection{Supply-chain replanning metrics}
For the supplier-loss setting, the following metrics quantify footprint and cost:

\paragraph{Overage cost.}
\begin{equation}
O = \sum_{(i,j)\in E,\,k\in K} \alpha_{ij} c_{ijk} y_{ijk}^{o}
+ \sum_{i\in V,\,k\in K} \beta_i e_{ik} p_{ik}^{o},
\end{equation}
where $y_{ijk}^{o}$ and $p_{ik}^{o}$ denote over-capacity transportation and production, respectively.

\paragraph{Network changes.}
\begin{equation}
N_c = \big|\{a_i \mid p_i \neq p_i'\}\big|
+ \big|\{(i,j) \mid y_{ij}\neq y_{ij}'\}\big|.
\end{equation}

\paragraph{Network additions.}
\begin{equation}
N_a = \sum_{i\in V} \max\{0,\xi_i'-\xi_i\}
+ \sum_{(i,j)\in E} \max\{0,\zeta_{ij}'-\zeta_{ij}\},
\end{equation}
where $\xi_i,\zeta_{ij}\in\{0,1\}$ indicate whether an agent or edge is used in the nominal plan and $\xi_i',\zeta_{ij}'$ are the corresponding indicators after replanning.

\paragraph{Communication.}
For supply-chain centralized coordination,
\begin{equation}
M = 1 + 2|V| + N_c + N_a,
\end{equation}
which accounts for one global replan trigger, information requests and responses with all agents, and update notifications to changed agents and edges. For distributed coordination, $M$ counts all local request, response, award/selection, and confirmation exchanges used to construct the new plan.

\subsection{Supply-chain stochastic robustness metrics}
For the lead-time disruption setting, let $\hat{y}_{zjk}$ be the replanned flow from supplier $a_z$ to demand agent $a_j$ for product $k$, and let $v_{zjk,i}$ be its realized arrival time in replication $i$. Flow lateness is
\begin{equation}
\Delta_{zjk,i}=\max\{v_{zjk,i}-t_{jk},0\}.
\end{equation}
The delayed-product percentage at a delay level $\delta_i$ is the fraction of total disrupted-demand flow whose realized lateness equals $\delta_i$:
\begin{equation}
\frac{\sum_{a_j\in A_{\mathrm{dm}},\,a_z\in Z_j(k),\,k\in K_d}
\hat{y}_{zjk}\;\mathbf{1}[\Delta_{zjk,i}=\delta_i]}
{\sum_{a_j\in A_{\mathrm{dm}},\,a_z\in Z_j(k),\,k\in K_d} \hat{y}_{zjk}}.
\end{equation}
At the aggregate demand-agent level,
\begin{equation}
C_{\mathrm{dm}} = \sum_{a_j\in A_{\mathrm{dm}}} C_{d,j},
\qquad
L_{\mathrm{dm}} = \sum_{a_j\in A_{\mathrm{dm}}}\sum_{k\in K_d} \Delta t_{jk},
\end{equation}
and the deterministic plan objective is
\begin{equation}
J_{\mathrm{dm}} = C_{\mathrm{dm}} + w_t L_{\mathrm{dm}},
\end{equation}
with lateness penalty weight $w_t=10^5$ in the reported case study.

\section{Unified Diagnostics Construction}
\label{app:diagnostics}

\subsection{R1: quality--latency--communication frontier}
R1 combines communication $M$, latency $T$, and a domain-specific quality signal into a single cross-domain diagnostic. In the manufacturing domain, the plotted frontier uses cycle time as the primary throughput-quality proxy together with average communication and replanning runtime. In the supply-chain domain, the plotted frontier uses scenario-level overage cost $O$ against communication $M$, with feasibility additionally marked as \emph{feasible} or \emph{unmet}.

\subsection{R2: propagation footprint and network structure}
R2 is reported on the supply-chain network-structured setting. The propagation footprint is represented by the footprint proxies $(N_c,N_a)$ together with communication $M$ and quality impact $O$. In the AIMS presentation, scenarios are stratified into four groups based on low/high connectivity and low/high redundancy, denoted LL, LH, HL, and HH. The current grouped analysis contains $53$, $16$, $2$, and $13$ scenarios in these four groups, respectively.

\subsection{R3: robustness under uncertainty}
R3 combines the two stochastic settings. In manufacturing, robustness is captured by incident counts (damaged products, broken machines, rescheduled processes) and risk summaries (peak and average schedule risk). In supply chain, robustness is captured by the delayed-product distribution across disruption intensities and by deterministic cost--lateness trade-offs summarized through $(C_{\mathrm{dm}},L_{\mathrm{dm}},J_{\mathrm{dm}})$.

\subsection{Inclusion criteria}
For the supplier-loss study, not all disruptions admit feasible recovery under local or centralized coordination. Cases with zero redundancy or insufficient residual capacity can become unmet-demand scenarios. The attribute-stratified propagation analysis is therefore restricted to the $73$ scenarios in which both centralized and distributed recovery produce feasible plans. The stochastic lead-time analysis is performed on the fixed disrupted-subnetwork centered on the cluster supplier $S3$ and the three affected cockpit assemblers.

\section{Experimental Protocol, Budgets, and Accounting}
\label{app:protocol}

\subsection{Trials, scenarios, and replications}
The manufacturing study reports averages over five simulation trials. The supplier-loss supply-chain study evaluates $84$ single-supplier-loss scenarios, of which $73$ are used for feasible-plan comparison and attribute-stratified diagnostics. The stochastic lead-time study evaluates three disruption intensities ($20\%$, $60\%$, $100\%$ lead-time increase) and uses $300$ out-of-sample simulation replications for each setting.

\subsection{Platform and solver notes}
The manufacturing environment is implemented in Repast Symphony. The stochastic supply-chain simulation is implemented using Python scripts that schedule agent decisions and simulate the resulting discrete-event dynamics. Supplier-response and supplier-selection problems are solved as tractable mixed-integer models under sample-average approximation in the stochastic setting.

\subsection{Budget definitions}
The mechanism is evaluated under explicit communication and time budgets. Communication is bounded by a message budget $B_{\mathrm{msg}}$ (and optionally a round budget), while latency is constrained by a wall-clock deadline $B_{\mathrm{time}}$. In the presented diagnostics, latency is reported as measured runtime and communication is reported as message exchanges under a fixed accounting rule.

\subsection{Message accounting rule}
For the appendix calculations and the unified diagnostics, each point-to-point \textsc{request}, \textsc{offer}, \textsc{award}, and \textsc{confirm} transmission is counted as one message exchange. If a message is sent to multiple recipients, the accounting is per recipient. Centralized coordination is accounted for separately using a domain-specific aggregation of information-collection and update-notification steps; distributed coordination counts only the explicitly executed negotiation messages needed to produce the revised plan.

\section{Additional Algorithmic and Qualitative Details}
\label{app:additional}

\subsection{Manufacturing fallback behavior}
If no feasible replacement sequence is found within the local coordination space---for example because no redundant resource is available or hard requirements cannot be met---the distributed manufacturing controller escalates to a central controller or human manager for further action or for relaxation of constraints. This fallback is not the default path; it is only used when the local candidate set is empty.

\subsection{Supply-chain propagation semantics}
Supply-chain communication proceeds in contract-style rounds. A disrupted agent informs directly affected downstream agents of the lost or delayed flow. These agents become demand agents and request support from candidate upstream suppliers. Once a demand agent selects replacement flows, the selected suppliers update their own local balance; if those commitments create new upstream needs, those suppliers become new demand agents. Propagation stops when all induced needs are balanced or when no additional feasible support exists.

\subsection{Qualitative effect of heterogeneous risk attitudes}
In the lead-time disruption case, heterogeneous risk attitudes change which backup suppliers are chosen even when all affected agents face the same disrupted supplier. In the reported examples, risk-averse decisions place less reliance on the disrupted supplier $S3$ and shift more quantity toward backup suppliers with better tail behavior, even when that choice incurs a modest increase in nominal cost. This effect is strongest at higher disruption intensity, where conservative commitments reduce severe delay outcomes.

\section{Scope and Remaining Limitations}
\label{app:limitations}

The current appendix closes several implementation gaps, but it does not claim a fully standardized industrial benchmark suite. The presented evidence covers two representative industrial domains and two supply-chain disruption regimes, but not a full cross-domain benchmark with a single shared simulator and protocol.

The mechanism exposes explicit scope and gate knobs---notably $(h_s,K_s)$ and $(\varepsilon_{\Delta},\varepsilon_r)$---that are important for deployment. The current paper reports their functional role and the resulting trends, but does not provide a systematic sensitivity sweep or automated tuning procedure.

Communication is accounted for at the message-exchange level rather than at the bit or semantic-content level. This is appropriate for the present mechanism-level comparison, but it does not capture compression, variable payload size, or network-layer contention.

Finally, the present stochastic supply-chain analysis focuses on a single disrupted supplier with heterogeneous risk attitudes primarily at the demand side. More diverse agent-risk combinations, simultaneous disruptions, and broader topological regimes remain open extensions.


\end{document}